\documentclass[11pt]{article}
\usepackage{aas_macros,amsmath,amssymb,comment,cite,esint,graphicx,mathtools,diagbox}
\usepackage{bm}
\usepackage[margin=.8in,letterpaper]{geometry}
\usepackage[colorlinks=true]{hyperref}
\usepackage[affil-it]{authblk}
\usepackage{subcaption}
\usepackage[utf8]{inputenc}
\usepackage{mathrsfs}
\usepackage{appendix}
\usepackage{amssymb}
\usepackage{float}                  
\usepackage{color}
\usepackage{cite}
\usepackage{hyperref}
\hypersetup{pageanchor=false}
\usepackage{indentfirst}
\usepackage{url}
\usepackage{xfrac}
\usepackage{caption}
\usepackage[numbers,square,comma,sort&compress,merge]{natbib}
\usepackage{esint}
\usepackage{overpic}
\usepackage{graphicx}
\usepackage{epsf,amsmath,bbold,amsfonts,stmaryrd}
\usepackage{textcomp}
\usepackage{ulem}
\usepackage{tikz}
\usepackage{multirow}
\numberwithin{equation}{section}
\let\originalleft\left
\let\originalright\right
\renewcommand{\left}{\mathopen{}\mathclose\bgroup\originalleft}
\renewcommand{\right}{\aftergroup\egroup\originalright}

\def\bea{\begin{eqnarray}}
\def\eea{\end{eqnarray}}
\def\nn{\nonumber}

\usepackage{tensor}
\usepackage{physics}

\newcolumntype{P}[1]{>{\Centering\hspace{0pt}}p{#1}}
\newcolumntype{Z}{>{\centering\arraybackslash}X} 

\newcommand{\df}{\mathrm{d}}   
   
\newcommand{\pa}[1]{\left(#1\right)}

\newcommand{\dmt}{{\delta_\text{mt}}}
\newcommand{\dmp}{{\delta_\text{mp}}}

\newcommand{\rt}{{r_\text{t}}}
\newcommand{\pt}{{\phi_\text{t}}}
\newcommand{\rmt}{{r_\text{mt}}}
\newcommand{\pmt}{{\phi_\text{mt}}}
\newcommand{\rmp}{{r_\text{mp}}}
\newcommand{\pmp}{{\phi_\text{mp}}}
\newcommand{\rhotm}{{\rho_\text{tm}}}
\newcommand{\vtm}{{\varphi_\text{tm}}}
\newcommand{\rhot}{{\rho_\text{t}}}
\newcommand{\vt}{{\varphi_\text{t}}}
\newcommand{\rhop}{{\rho_\text{p}}}
\newcommand{\vp}{{\varphi_\text{p}}}

\begin{document}
\title{\bf Semi-analytical Study on the Polarized Images of Black Hole due to Frame Dragging }
	
\author{Xinyu Wang$^{1,2}$, Songbai Chen$^{3,4}$, Minyong Guo$^{1,2\ast}$, Bin Chen$^{5,6}$}
\date{}
	
\maketitle
\vspace{-15mm}

\begin{center}
{\it
$^1$ School of physics and astronomy, Beijing Normal University,
Beijing 100875, P. R. China\\\vspace{2mm}

$^2$Key Laboratory of Multiscale Spin Physics (Ministry of Education), Beijing Normal University, Beijing 100875,  P. R. China\\\vspace{2mm}

$^3$ Department of Physics, Institute of Interdisciplinary Studies, Hunan Research Center of the Basic
Discipline for Quantum Effects and Quantum Technologies, Key Laboratory of Low Dimensional
Quantum Structures and Quantum Control of Ministry of Education, Synergetic Innovation Center
for Quantum Effects and Applications, Hunan Normal University, Changsha, Hunan 410081,  P. R. China\\\vspace{2mm}

$^4$ Center for Gravitation and Cosmology, College of Physical Science and Technology,
Yangzhou University, Yangzhou 225009, P. R. China\\\vspace{2mm}

$^5$ Institute of Fundamental Physics and Quantum Technology, \\
\& School of Physical Science and Technology,\\Ningbo University, Ningbo, Zhejiang 315211, P. R. China\\\vspace{2mm}

$^6$ School of Physics, Peking University, \& Center for High Energy Physics, \\No.5 Yiheyuan Rd, Beijing 100871, P.R. China\\\vspace{2mm}
}
\end{center}

\vspace{8mm}

\begin{abstract}

An initially retrograde accretion flow is transformed into a prograde configuration before plunging into the black hole, as a result of the frame-dragging effect induced by the black hole’s rotation. The polarized image of a black hole shaped by such an accretion flow manifests three distinctive critical locations: the turning point of the flow’s primary image, the polarization-flip location on the image plane, and the position of the primary image corresponding to the flow’s actual turning point in spacetime. Due to the influences of gravitational lensing and gravitational Faraday rotation, these three positions generally do not coincide. In this work, we examine a thin equatorial accretion disk composed of initially retrograde, geodesically moving flows, and conduct a systematic investigation into the interrelations and discrepancies among these critical locations. We elucidate the spatial hierarchy among the three, and for an on-axis observer, we derive approximate analytic expressions characterizing their positions.

\end{abstract}

\vfill{\footnotesize $\ast$ Corresponding author: minyongguo@bnu.edu.cn}

\maketitle

\newpage
\baselineskip 18pt
\section{Introduction}\label{sec1}

In general relativity, the rotation of a massive body distorts the surrounding spacetime, producing a gravitomagnetic frame-dragging effect—the Lense–Thirring precession \cite{1918PhyZ...19..156L}—which has been confirmed through weak-field experiments \cite{2004Natur.431..958C,Everitt:2011hp}. Moreover, in the vicinity of a rotating black hole’s event horizon, the phenomenon of frame dragging becomes exceedingly pronounced, compelling all matter and radiation to co-rotate with the black hole's intrinsic angular momentum \cite{Johnson:2019ljv,Wang:2023nwd,Wang:2024uda}. The existence of black holes in the universe is now supported by a wealth of evidence, ranging from the detection of gravitational waves by LIGO/Virgo \cite{LIGOScientific:2016aoc} to the groundbreaking images captured by the Event Horizon Telescope (EHT) \cite{EventHorizonTelescope:2019dse,EventHorizonTelescope:2022wkp}. Observations from the EHT \cite{EventHorizonTelescope:2021bee,EventHorizonTelescope:2024hpu} have also provided polarization measurements, offering deeper insights into the astrophysical environments in the immediate vicinity of black holes. Thus, utilizing polarized images of black holes has made it possible to investigate the frame-dragging effect caused by their rotation.

At its core, a polarized image of a black hole represents the synchrotron emission produced by electrons within the plasma surrounding the black hole, as projected onto the observer’s plane by the gravitational lensing effects of the black hole itself \cite{Ricarte:2022sxg,Chael:2023pwp,Zhang:2023cuw,Chen:2022scf, Huang:2024gtu}. These polarized images contain crucial information not only about the distribution of intensity but also about linear polarization, the latter being characterized by the electric vector position angle (EVPA) \cite{1994ApJ...427..718R,Himwich:2020msm,Vincent:2023sbw,Chen:2024jkm,Gelles:2021kti,Hou:2024qqo,Yang:2025byw,Qin:2022kaf,Chen:2024cxi}. Owing to the presence of the frame-dragging effect around a rotating black hole, an accretion flow that is initially retrograde may undergo a transition to prograde motion as it approaches sufficiently close to the black hole \cite{Ricarte:2022wpd}. For polarized black hole images formed by such accretion flows, two distinct observational signatures can be attributed to the influence of frame dragging.

One is the presence of a turning point in the primary image on the observer’s screen; the other is the existence of a polarization-flip location in the EVPA \cite{Ricarte:2022wpd}. The former arises due to the influence of frame dragging, under which the accretion flow transitions from retrograde to prograde motion, thereby imparting a corresponding turning point in its primary image. The latter originates from the frame-dragging effect and the magnetic flux-freezing effect in ideal magnetohydrodynamics, wherein the magnetic field is aligned with the fluid’s motion in the comoving frame \cite{PhysRevD.12.2959,Hou:2023bep}. Since the polarization is predominantly perpendicular to the magnetic field, a reversal in the direction of the accretion flow induces a polarization-flip phenomenon \cite{Chael:2023pwp,Palumbo:2020flt,Ricarte:2022wpd}. Related phenomena have been investigated in earlier works, in which general relativistic numerical simulations \cite{Ricarte:2022wpd} were performed to identify strong-field signatures of frame dragging in cases where infalling gas reverses its rotation during accretion, and hotspot models were analyzed analytically \cite{Chen:2024jkm} to characterize the change in the spiral orientation of polarization lines induced by frame dragging.

However, when accounting for the effects of gravitational lensing \cite{Gralla:2019drh} and gravitational Faraday rotation \cite{1977ApJ...212..541P,1982GReGr..14..865F,Brodutch:2011qt,Ishihara:1987dv,Gelles:2021kti} on polarization, the turning point of the primary image of the streamline and the location of the polarization flip generally do not coincide. Yet, in previous studies, the spatial relationship between these two features, as well as the extent of their discrepancy, has not been systematically investigated. Moreover, although both phenomena are linked to the reversal of the accretion flow, the position of the primary image corresponding to the actual flow reversal remains unknown in relation to the other two. Therefore, in this work, we undertake a systematic exploration of the spatial correlations and differences among these three critical locations. Specifically, we assume that the thin accretion disk on the equatorial plane is composed of accretion flows initially in retrograde motion along geodesic trajectories. We then compute the locations of the turning point of the primary image of the streamline, $(\rhot, \vt)$; the polarization-flip location, $(\rhop, \vp)$; and the position of the primary image corresponding to the actual flow reversal, $(\rhotm, \vtm)$. By examining the discrepancies among these three positions under varying viewing angles and flow-reversal radii, we discover that the ordering $\rhotm < \rhop < \rhot$ consistently holds for the same azimuthal angle, and that the separation between $\rhop$ and $\rhot$ is significantly smaller than the separations between either of them and $\rhotm$. Specifically, for on-axis observers, we analytically examine the relationship among the three under a certain approximation.

The remaining sections of this paper are structured as follows. In Sec. \ref{sec2}, we present the essential concepts of black hole spacetimes, the accretion flow model considered in this work, and the methodology for computing polarization. Sec. \ref{sec3} provides a semi-analytical analysis for the on-axis case, focusing on the relation among the three characteristic locations and extends the study numerically to off-axis observers, quantifying the dependence of these locations on the turning radius $\rt$  and the inclination angle $\theta_o$. Finally, in Sec. \ref{sec4} we provide a summary and outlook. In this work, we have set the fundamental constants $c$, $G$ to unity, and we will work in the convention $(-,+,+,+)$.

\section{Preliminary and method}\label{sec2}
In this section, we present the essential concepts related to black hole metrics, timelike and null geodesics, the accretion flow model under examination, and the methodologies employed for polarization studies.

\subsection{Kerr black hole and geodesics}

The metric of the Kerr black hole in Boyer-Lindquist (BL) coordinates can be expressed as  
\bea\label{metric}  
\df s^2 = -\left(1 - \frac{2 M r}{\Sigma}\right) \df t^2 + \frac{\Sigma}{\Delta} \df r^2 + \Sigma \df \theta^2 + \left(r^2 + a^2 + \frac{2 M r a^2}{\Sigma} \sin^2 \theta\right) \df \phi^2 - \frac{4 M r a}{\Sigma} \sin^2 \theta \df t \df \phi\,,  
\eea  
where  
\bea  
\Delta = r^2 - 2 M r + a^2\,, \quad \Sigma = r^2 + a^2 \cos^2 \theta\,.  
\eea  
\(M\) and \(a\) are the mass and spin parameters of the Kerr black hole, respectively. Furthermore, for the sake of simplicity and without loss of generality, we shall set \(M = 1\) in the forthcoming discussion. The horizons are \(r_{\pm} = 1 \pm \sqrt{1 - a^2}\), which represent the roots of the equation \(\Delta = 0\).  
The geodesic equations are completely integrable since they admit four constants along the trajectory of a particle: \(g_{\mu\nu}u^\mu u^\nu = -\mu^2\), with \(\mu^2 = 0\) and \(\mu^2 = 1\) for null and timelike geodesics, respectively; the energy \(E = -u \cdot \partial_t\); the angular momentum \(L = u \cdot \partial_\phi\); and the Carter constant \(Q\). Here, we have assumed \(u^\mu = \frac{dx^\mu}{d\tau}\), where \(u^\mu\) is interpreted as the four-velocity or the four-momentum per unit mass for timelike geodesics, with \(\tau\) representing the proper time. The physical meanings of \(E\) and \(L\) in this context correspond to the energy and angular momentum per unit mass, respectively. For null geodesics, \(u^\mu\) denotes the photon's four-wavevector, and \(\tau\) should be understood as the affine parameter along the worldline. In this case, \(E\) and \(L\) are independent of mass.

By utilizing these conserved quantities, the geodesic equations can be written in first-order form \cite{Bardeen:1973tla}:  
\bea  
\label{fv}  
u^r & = & \pm_r \frac{1}{\Sigma} \sqrt{\mathcal{R}(r)} \,, \nn\\  
u^\theta & = & \pm_\theta \frac{1}{\Sigma} \sqrt{\Theta(\theta)} \,, \nn\\  
u^\phi & = & \frac{1}{\Sigma} \left[\frac{a}{\Delta} \left(E\left(r^2 + a^2\right) - a L\right) + \frac{L}{\sin^2 \theta} - a E\right] \,, \nn\\  
u^t & = & \frac{1}{\Sigma} \left[\frac{r^2 + a^2}{\Delta} \left(E\left(r^2 + a^2\right) - a L\right) + a\left(L - a E \sin^2 \theta\right)\right] \,.  
\eea  
Here,  
\bea\label{pots}  
\mathcal{R}(r) & = & \left[E\left(r^2 + a^2\right) - a L\right]^2 - \Delta \left[Q + (L - a E)^2 + \mu^2 r^2\right] \,, \nn\\  
\Theta(\theta) & = & Q + a^2\left(E^2 - \mu^2\right) \cos^2 \theta - L^2 \cot^2 \theta \,,  
\eea  
are the radial and angular potentials, respectively. \(\pm_r\) and \(\pm_\theta\) denote the signs of the radial and polar motions.

\subsection{The accretion disk model}

In this analysis, for the sake of simplicity, we consider an analytic thin accretion disk model situated on the equatorial plane. The accretion flow under examination is assumed to exhibit symmetry about the \(z\)-axis, which corresponds to the rotational axis of the Kerr black hole. Specifically, we commence by selecting a timelike plunging trajectory confined to the equatorial plane, with the form of the four-velocity given as 
\[
u^\mu = (\dot{t}, \dot{r}, \dot{\theta} = 0, \dot{\phi}) \big|_{\theta = \pi / 2}\,.
\]

\begin{figure}[htbp]
    \centering
    \centering
    \includegraphics[width=3in]{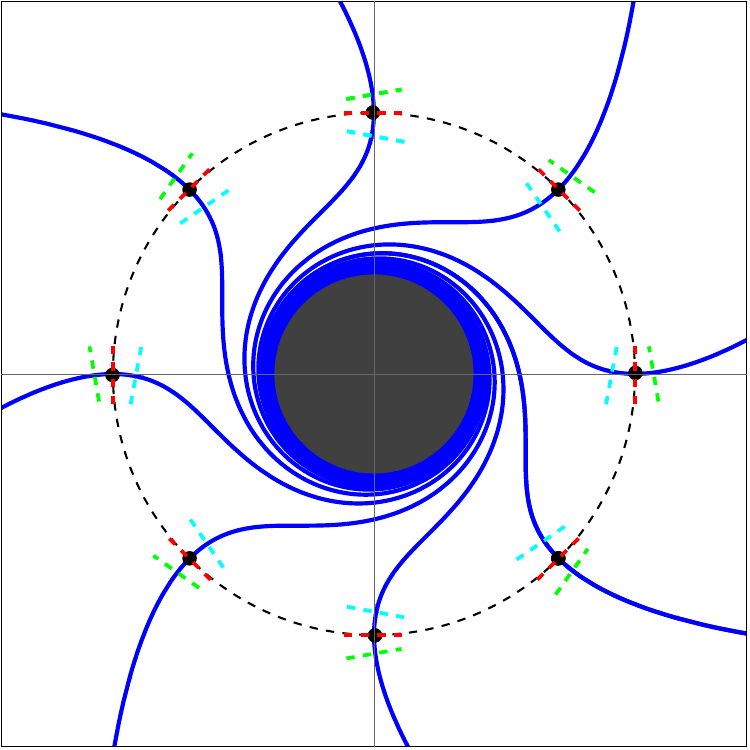}
    \caption{The schematic diagram of the accretion disk is presented on the equatorial plane. The central black region represents the black hole. The blue curves illustrate the accretion streamlines of the disk, which are symmetric with respect to the origin. The position of the black dot corresponds to the location where the accretion flow's \( u^{\phi} = 0 \,\).}
    \label{geos}
\end{figure}

This trajectory is then rotated around the $z$-axis, constructing a model of the thin accretion disk. Consequently, the four-velocity at every point on the accretion disk is fully determined. In Fig.~\ref{geos}, we present a schematic diagram of the accretion disk model under consideration.Without loss of generality, we have illustrated eight accretion streamlines that are symmetric about the origin. We first draw one such streamline, then generate the remaining seven trajectories by rotating it at intervals of $\pi/4$. The accretion streamlines we consider plunge from outside, and for large $r$, $u^{\phi} < 0 \,;$ under the influence of the black hole's frame-dragging effect, $u^{\phi}$ eventually becomes greater than $0$. Thus, there exists a turning point where $u^{\phi} = 0 \,$, denoted by $\rt$, which we represent with a black dot in the figure. Evidently, due to the rotational symmetry of the accretion flow, these black dots are distributed along a circle.

\begin{figure}[h]
    \centering
    \includegraphics[width=5.2in]{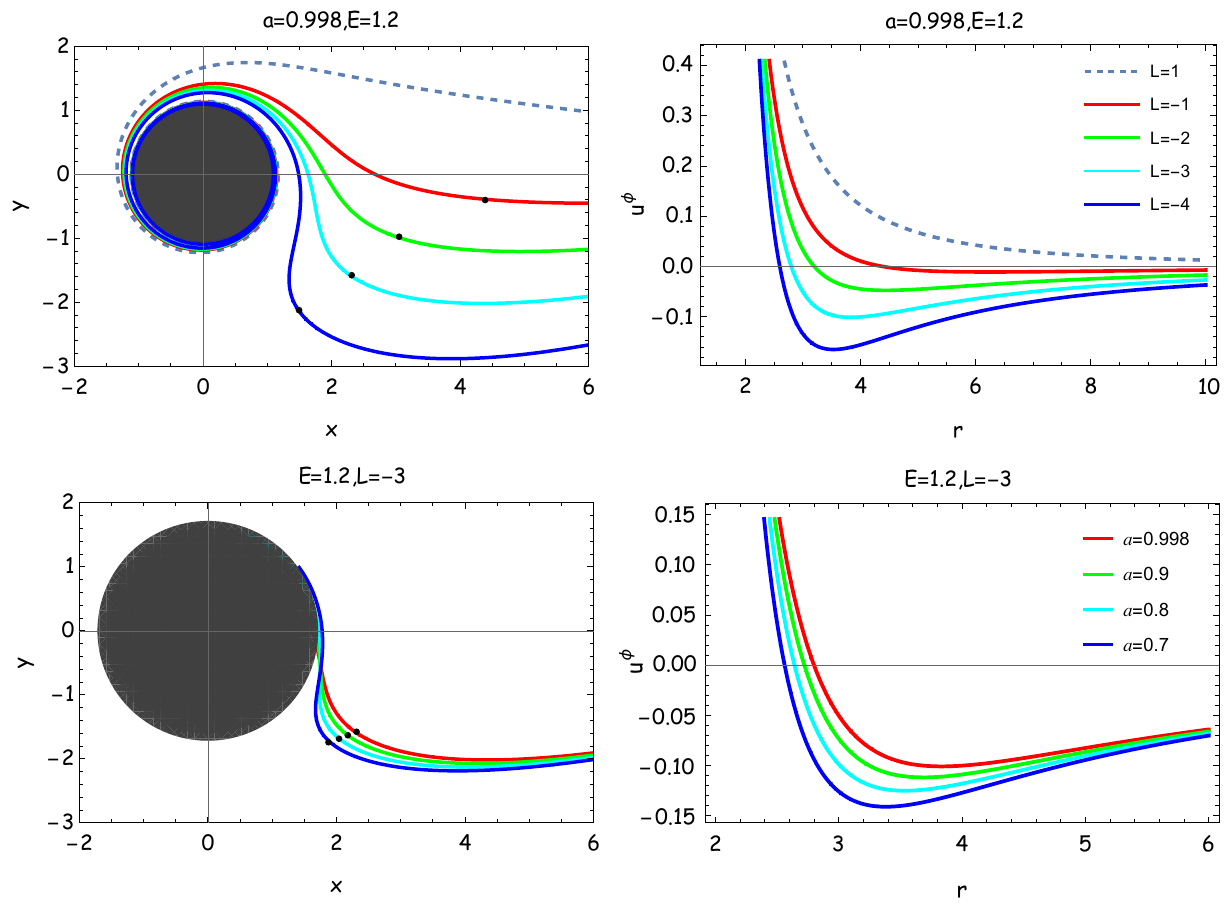}
    \caption{Top left panel: Trajectories of geodesics plunging from \(r=15\) in the \(xy\)-coordinate system, with \(x=r\cos\phi\) and \(y=r\sin\phi\). The dashed, red, green, cyan, and blue lines correspond to angular momentum values \(L=1\), \(L=-1\), \(L=-2\), \(L=-3\), and \(L=-4\), respectively. The energy and spin parameters are fixed at \(E=1.2\) and \(a=0.998\).
Top right panel: The azimuthal velocity \(u^\phi\) is illustrated as a function of the radial coordinate \(r\). The parameters used match those in the top left panel.
Bottom left panel: Trajectories of geodesics plunging from \(r=15\) in the \(xy\)-coordinate system. The red, green, cyan, and blue lines represent spin parameters \(a=0.998\), \(a=0.9\), \(a=0.8\), and \(a=0.7\), respectively, with energy fixed at \(E=1.2\) and angular momentum \(L=-3\).
Bottom right panel: The azimuthal velocity \(u_\phi\) is plotted as a function of the radial coordinate \(r\). The parameters correspond to those used in the bottom left panel. Black dots indicate the inversion points where \(u^\phi=0\). The shaded gray region denotes the interior of the event horizon.}
    \label{fig1}
\end{figure}

For the specific accretion flow streamlines, in this work we consider the motion of the accretion flow as dictated by the geodesic equation. Furthermore, in consideration of the Kip Thorne limit for rotating black holes \cite{1974ApJ...191..507T}, we assume the spin parameter $a$ of the black hole reaches its maximum value of $0.998$. From the radial potential $\mathcal{R}(r) \geq 0$ in Eq. (\ref{pots}), it is evident that timelike geodesic particles at infinity satisfy $E \geq 1$. Thus, in the case of accretion flows governed by geodesic motion, we adopt $E = 1.2$. As this work primarily focuses on black hole images induced by frame dragging, we concentrate on accretion streamlines that include the condition $u^\phi = 0$. From Eq. (\ref{fv}), it is equivalent to the existence of a turning point $\rt$ in the azimuthal direction in the BL coordinates satisfying
\bea
\label{rtb}
\rt=2-\frac{2 a E}{L}>2>r_h\,,
\eea
where $r_h=r_+$ is the radius of the event horizon. Based on these considerations, within the geodesic accretion flow model, for a specific example, we fix $E = 1.2$ and $a = 0.998$, allowing $L$ to vary as $L = -1 \, , -2 \, , -3 \, , -4$. Additionally, we include a comparison case for $L = 1$, wherein $\rt$ does not exist. Moreover, we fix $E = 1.2$ and $L = -3$, and explore the variation of $a$ by setting $a = 0.998 \, , 0.9 \, , 0.8 \, , 0.7$. The variations in the timelike trajectories and $u^{\phi}$ for these scenarios are depicted in Fig. \ref{fig1}. From the figure, it is readily apparent that for the trajectory with $L=1$, $u^\phi$ outside the horizon lacks a turning point, whereas for the other trajectories, a corresponding $u^\phi = 0$ turning point exists outside the horizon.

\subsection{The magnetic field and polarization of synchrotron radiation}

In this subsection, we introduce the magnetic configuration of interest in the vicinity of a Kerr black hole. First, we assume the ideal MHD condition, expressed as
\begin{equation}
F_{\mu\nu} u^\nu = 0 \, .
\end{equation}
Given the axisymmetric and stationary nature of the accretion flow, we assume that the gauge potential of the magnetic field is independent of both time and azimuthal coordinates, i.e., $A_\mu = A_\mu(r, \theta)$. Furthermore, by considering Maxwell's equations, we obtain the following expression \cite{PhysRevD.12.2959,Hou:2023bep}:
\begin{equation}
\label{mag}
B^\mu = \frac{1}{\sqrt{-\det(g_{\mu\nu})}} \frac{\partial_\theta A_\phi}{u^r} \left[\left(u_t + \Omega_B u_\phi\right) u^\mu + \Omega_B \delta^\mu_\phi + \delta^\mu_t\right], \quad \mu = t, r, \theta, \phi,
\end{equation}
where $\det(g_{\mu\nu}) = -\Sigma^2 \sin^2\theta$, and $\Omega_B$ is the field line angular velocity, which characterizes the rotation of magnetic field lines. In this work, we do not concern ourselves with the influence of $\Omega_B$, and its specific value bears no substantive impact on our analysis; hence, without loss of generality, we set $\Omega_B = 0$ in the discussions that follow.  From Eq.\,(\ref{mag}), it is evident that the spatial components 
\(B^i\) are parallel to \(u^i\), indicating that the magnetic field is effectively frozen into the streamlines. The function $\partial_\theta A_\phi$ can be chosen to adopt a split monopole configuration, as proposed by \cite{Blandford:1977ds}:
\begin{equation}
\partial_\theta A_\phi = \Phi_0 \, \mathrm{sign}(\cos\theta) \sin\theta \, ,
\end{equation}
where $\Phi_0$ is a constant. 

Due to the presence of a magnetic field, rapidly moving thermal or nonthermal electrons within the accreting matter undergo synchrotron radiation as a result of the Lorentz force. The polarization vector \( f^\mu \) is predominantly perpendicular to the global magnetic field, leading to the expression \cite{Hou:2024qqo}
\begin{equation}\label{defpv}
f^\mu = \frac{\epsilon^{\mu\nu\alpha\beta} u_\nu k_\alpha B_\beta}{\sqrt{|f^2|}} \, ,
\end{equation}
where \( \epsilon^{\mu\nu\alpha\beta} \) denotes the Levi-Civita tensor and the quantity \( k^\mu \) signifies the photon’s four-momentum, while \( B^\mu \) describes the magnetic field in the Boyer-Lindquist coordinate basis. The unit-norm polarization vector \( f^\mu \) is orthogonal to the photon’s four-momentum, so that  
\begin{equation}  
f^\mu k_\mu = 0 \, , \quad f^\mu f_\mu = 1\,.
\end{equation}

With this, the essential discussion on the light source concludes. Let us briefly summarize: We have considered a geometrically thin accretion disk in the equatorial plane as the radiation source, constructed by first prescribing a plunging accretion flow and then generating it through rotation. Additionally, we have incorporated the magnetic field structure accompanying the accreting material. Electrons within the accretion flow are accelerated by the magnetic field, resulting in synchrotron radiation and the subsequent production of polarized light. Furthermore, we have specified the polarization vector at the source and the equations it satisfies.

\subsection{Numerical ray-tracing and radiative transfer}

In this subsection, we introduce the ray-tracing, radiative transfer and the decomposition of linear polarization employed in the present work. Regarding the ray-tracing, we closely follow the scheme used in our previous work \cite{Hu:2020usx, Hou:2022eev, Zhong:2021mty}, which has been applied to many interesting studies \cite{Li:2025awg,Zhen:2025nah,Liu:2025lwj}. We use the zero-angular-momentum observer (ZAMO) \cite{Bardeen:1973tla,Frolov:1998wf,Cunha:2016bpi} to model the distant observer, and the tetrad is given by  
\bea  
e_{(0)}&=\xi(1,0,0,-\gamma)\,, \quad e_{(1)}=\left(0,-\frac{1}{\sqrt{g_{r r}}}, 0,0\right)\,, \\  
e_{(2)}&=\left(0,0, \frac{1}{\sqrt{g_{\theta \theta}}}, 0\right)\,, \quad e_{(3)}=\left(0,0,0,-\frac{1}{\sqrt{g_{\phi \phi}}}\right)\,,  
\eea  
where  
\bea  
\xi=\sqrt{\frac{-g_{\phi \phi}}{g_{t t} g_{\phi \phi}-g_{t \phi}^2}}\,, \quad \gamma=\frac{g_{t \phi}}{g_{\phi \phi}}\,.
\eea

The four-momentum of photons in the ZAMO coordinate system can be expressed as:  
\bea  
p_{(\mu)} = k_\nu e_{(\mu)}^\nu \,,  
\eea  
where \( k^\mu \) represents the photon's four-momentum. By utilizing celestial coordinates, we can determine the projections onto the observer’s screen. To achieve this, we introduce a Cartesian coordinate system on a square observation screen. In this system, the $X$-axis is aligned with \( -e_{(3)} \), the $Y$-axis is aligned with \( -e_{(2)} \), and the origin is shifted along \( -e_{(1)} \) in the ZAMO frame by an amount corresponding to the photon energy measured by the observer. 

For image construction on the observation screen, we employ the stereographic projection method, as outlined in our previous work \cite{Hu:2020usx}. In this method, the relationship between the Cartesian coordinates \( (X, Y) \) and the four-momentum is established using the celestial coordinates \( \Theta \) and \( \Psi \). These celestial coordinates can be expressed in terms of the four-momentum components as follows:  
\bea  
\cos \Theta = \frac{p^{(1)}}{p^{(0)}} \,, \quad \tan \Psi = \frac{p^{(3)}}{p^{(2)}} \,.  
\eea  
Additionally, the Cartesian coordinates \( (X, Y) \) on the observation screen are given by:  
\bea  
X = -2 \tan \frac{\Theta}{2} \sin \Psi \,, \quad Y = -2 \tan \frac{\Theta}{2} \cos \Psi \,.  
\eea

Next, we direct our focus to the transfer of radiation. In general, the treatment of radiation transfer in curved spacetime is highly intricate, particularly when incorporating the evolution of polarization. However, in this work, we confine our attention to a geometrically thin accretion disk in Kerr spacetime, which substantially simplifies the analysis. Disregarding the influence of the geometrically thin disk on polarized radiation transfer, we separate the radiative transfer of intensity from that of polarization. For the former, we employ the approach utilized in our previous study \cite{Hou:2022eev}, which is concisely outlined as follows.

Assuming that refraction within the disk medium is negligible, the evolution of intensity adheres to the equation \cite{Lindquist:1966igj}:
\bea
\frac{d}{d \lambda} \left(\frac{I_\nu}{\nu^3} \right) = \frac{J_\nu - \kappa_\nu I_\nu}{\nu^2} \,,
\eea
where 
$\lambda$ denotes the affine parameter along null geodesics, and 
$I_\nu$, 
$J_\nu$, and 
$\kappa_\nu$ represent the specific intensity, emissivity, and absorption coefficient at frequency $\nu$, respectively. In a vacuum, both $J_\nu$ and $\kappa_\nu$ vanish, ensuring the conservation of $I_\nu / \nu^3$ along geodesics. Assuming that the accretion disk is geometrically thin, the radiation and absorption coefficients can be approximated as constants while the light propagates through the disk. We also assume that the accretion disk is optically thin, implying that absorption is negligible. Consequently, the intensity corresponding to a pixel on the observer’s screen can be computed as follows \cite{Hou:2022eev}:

\bea
I_{\nu_o} = \sum_{n=1}^{N_{\max}} g_n^3 J_n \,.
\eea
Here, \( \nu_o \) represents the observed frequency on the screen,  \( \nu_n \) is the frequency
measured in the local co-moving frame with respect to the accretion flow, \( n = 1, \dots, N_{\max} \) denotes the number of times the ray intersects the disk, with \( N_{\max} \) being the maximum number of crossings, \( g_n=\nu_o/\nu_n \) corresponds to the associated redshift factor, and \( J_n \) signifies the emissivity at the \( n \)-th intersection point. The emissivity is chosen as
\bea
J = \exp \left(-\frac{1}{2} z^2 - 2z \right) \,, \quad z = \log \frac{r}{r_h} \,,
\eea
with \( r_h \) being the horizon radius on the equatorial plane.

For the transfer of polarization in Kerr spacetime, the polarization information can be transmitted from the source to the observer using the parallel transport equation,  
\bea \label{pteq} 
k^\mu \nabla_\mu f^\nu = 0 \,.  
\eea
At the observer's location, we have established the ZAMO frame along with the corresponding imaging screen. In our plane, the two orthogonal basis vectors of the screen plane are given by  
\begin{align}  
\hat{e}_X = -\hat{e}_{(3)} = \frac{\partial_\phi}{\sqrt{g_{\phi\phi}}} \,,\qquad  
\hat{e}_Y = -\hat{e}_{(2)} = -\frac{\partial_\theta}{\sqrt{g_{\theta\theta}}} \,.  
\end{align}  
Utilizing these basis vectors, we can obtain the projection of the polarization vector onto the coordinate plane as  
\begin{align}  
f_{X} = f^\mu \cdot \hat{e}_X \,,\qquad  
f_{Y} = f^\mu \cdot \hat{e}_Y \,.  
\end{align} 
 
Subsequently, we can transform these into the Stokes parameters \( Q \) and \( U \) for further computation. They are defined as follows:  
\begin{align}  
P = Q + iU\,,\qquad Q = \lvert P \rvert \pa{f_X^2 - f_Y^2} \,,  
\qquad  
U = 2\lvert P \rvert f_X f_Y \,.  
\label{eqn:rs}  
\end{align}  
where \( P \) is a complex-valued polarization field, whose modulus \( \lvert P \rvert \) represents the intensity of linearly polarized light. The expression for EVPA is given by  
\begin{align}  
\chi = \arctan \pa{\frac{f_Y}{f_X}}=\frac{1}{2} \arctan \pa{\frac{U}{Q}}=\frac{1}{2}\angle(P)\,,
\end{align}
where $\angle(P)$ is denotes complex phase of the polarization field. 

In the observer's two-dimensional screen, it is convenient to introduce a polar coordinate system expressed as
\bea
\frac{X}{r_o}=\rho\cos\varphi\,,\quad \frac{Y}{r_o}=\rho\sin\varphi\,,
\eea
where the radial distance \(\rho\) is measured from the image center and the azimuthal angle \(\varphi\) is measured counter-clockwise on our image. 

Then, we can determine the coordinates of the primary images $(\rhotm, \vtm)$ on the image plane from the turning points of the accretion flow $(\rt, \pt)$. Additionally, the turning point coordinates of the primary image can be identified directly on the image plane by solving $d\varphi = 0$; these are denoted as $(\rhot, \vt)$. Conversely, given the coordinates of the primary image’s turning point $(\rhot, \vt)$, we can trace back to infer the corresponding coordinates of the emitting region, denoted as $(\rmt, \pmt)$. It is important to note that, due to gravitational lensing effects, the coordinates $(\rhotm, \vtm)$ and $(\rhot, \vt)$, as well as the actual turning point of the accretion flow and the position $(\rmt, \pmt)$, generally do not coincide.

On the other hand, in light of magnetic field freezing effects, the presence of a turning point in the accretion stream implies a polarity flip in the magnetic field-induced polarization along the flow line. At this critical position, the EVPA becomes perpendicular to the radial direction on the screen, that is, when $f_\mu \cdot \hat{e}_\rho = 0$. We can denote the corresponding coordinates on the image plane as $(\rhop, \vp)$, with the associated source plane coordinates labeled $(\rmp, \pmp)$.

It is worth observing that, in the hypothetical absence of gravitational lensing and parallel transport of the polarization vector, the image and source planes become indistinguishable, causing all six aforementioned points to converge at a single location, as illustrated for the source plane in Fig.~\ref{geos}. In other words, by analysing the variations and discrepancies among these corresponding critical coordinates, we can extract richer physical information about both the black hole and the emitting source from the polarized image. Such insights are invaluable for deepening our understanding of black hole polarization images.

\subsection{Penrose–Walker constant}

In practical computations, the EVPA on the image plane is not directly obtained using Eq.~(\ref{pteq}). Instead, given that the Kerr spacetime belongs to the Petrov-type D classification, a conserved complex scalar, known as the Penrose–Walker (PW) constant, $\kappa$ \cite{Walker:1970un, Gelles:2021kti}, emerges. This invariant encapsulates the parallel transport of polarization vectors along null geodesics threading through the Kerr geometry. It is elegantly expressed as  
\begin{equation}
\begin{aligned}
\kappa &\equiv \kappa_1 + i \kappa_2 = (\mathcal{A} - i \mathcal{B})(r - i a \cos \theta), \\
\mathcal{A} &= (k^t f^r - k^r f^t) + a \sin^2\theta (k^r f^\phi - k^\phi f^r), \\
\mathcal{B} &= \left[ (r^2 + a^2)(k^\phi f^\theta - k^\theta f^\phi) - a (k^t f^\theta - k^\theta f^t) \right] \sin \theta,
\end{aligned}
\end{equation}
where $k^\mu$ denotes the photon's four-momentum, and $f^\mu$ signifies its associated polarization vector. From this complex scalar, the observable EVPA, $\chi$, can be extracted through the following relation:  
\begin{equation}
\chi = \arctan\left( \frac{\kappa_1}{\kappa_2} \right) + \frac{\pi}{2} + \varphi,
\end{equation}
where $\varphi$ represents the azimuthal angular coordinate on the observer’s image plane. Moreover, the critical condition $f_\mu \cdot \hat{e}_\rho = 0$ is equivalent to the relation $\chi - \varphi = \frac{\pi}{2}$, which in turn implies that $\kappa_1 = 0$. In particular, at the source location $(r_s,\theta_s = \frac{\pi}{2})$, the PW constant simplifies to  
\begin{equation}\label{abeq}
\begin{aligned}
\kappa & =\kappa_1+i \kappa_2=r_{\mathrm{s}}(\mathcal{A}-i \mathcal{B})\,, \\
\mathcal{A} & =\left(k^t f^r-k^r f^t\right)+a\left(k^r f^\phi-k^\phi f^r\right)\,, \\
\mathcal{B} & =\left[\left(r_{\mathrm{s}}^2+a^2\right)\left(k^\phi f^\theta-k^\theta f^\phi\right)-a\left(k^t f^\theta-k^\theta f^t\right)\right]\,.
\end{aligned}
\end{equation}  
so that the condition $f_\mu \cdot \hat{e}_\rho = 0$ is equivalent to $\mathcal{A} = 0$.

\subsection{On-axis observers}

The relationship between the source coordinates $(r, \phi)$ and the image plane coordinates $(\rho, \varphi)$ generally lacks an analytical expression. However, for an observer situated along the polar axis, $\theta_o = 0^\circ$, an approximate analytical formulation becomes attainable. From the angular potential appearing in the $\theta$-component of the geodesic equations, Eq.~(\ref{fv}), with $\mu^2 = 0$, we can see that only photons with zero angular momentum can reach the spin axis. When such a photon arrives at the screen of an on-axis observer situated at a large distance, its arrival position is described in polar coordinates as \cite{Gralla:2019drh, Gelles:2021kti, Chen:2024jkm}
\begin{equation}\label{reeq}
\rho = \sqrt{\eta + a^2}\,, \quad \varphi = \phi_o\,,
\end{equation}
where $\eta$ denotes the energy-rescaled Carter quantity $Q/E^2$, $r_o$ and $\phi_o$ are the radial and azimuthal coordinates of the arriving photon, respectively. 

For direct images, the relation between the emission point $(r_s, \phi_s)$ and the observed coordinates $(\rho, \varphi)$ can be well approximated by the following analytical expansions \cite{Gates:2020sdh, Gralla:2019drh, Gelles:2021kti, Chen:2024jkm}:
\bea\label{rbrr}
r_s &=& \rho - 1 + \frac{1 - a^2}{2 \rho} + \frac{3(5\pi - 16)}{4 \rho^2} + \mathcal{O}\left( \frac{1}{\rho^3} \right)\,, \nn\\
\rho &=& r_s + 1 + \frac{a^2 - 1}{2 r_s} + \frac{50 - 2 a^2 - 15 \pi}{4 r_s^2} + \mathcal{O}\left( \frac{1}{r_s^3} \right)\,.
\eea
Note that although the above equations were derived under the assumption that $r_s \gg M$, they remain remarkably accurate even near the horizon scale ($r_s \sim r_+$) across the full range of Kerr spin \cite{Gates:2020sdh, Gralla:2019drh, Gelles:2021kti, Chen:2024jkm}. Moreover, the arrival angle is determined by \cite{Chen:2024jkm}
\bea
\label{varphi}
\varphi = \phi_s + \int_{r_s}^{\infty} \frac{2a r \, dr}{\Delta \sqrt{\mathcal{R}}}, \quad \mathcal{R} = \left(r^2 + a^2\right)^2 - \Delta\left(\eta(r_s) + a^2\right),
\eea
where the integral in Eq.~\eqref{varphi} admits the approximate evaluation \cite{Chen:2024jkm}
\bea\label{phiaeq}
\varphi - \phi_s \approx \int_{r_s}^{\infty} \frac{2a \, dr}{r^2 \sqrt{r^2 - r_s^2}} = \frac{2a}{r_s^2},
\eea
thereby yielding a more concise expression.

\section{Results}\label{sec3}

In this section, we present the numerical results for the polarized images of Kerr black holes with thin accretion disks. Without loss of generality, we assume the black hole spin to be $a > 0$ in the following analysis.

\begin{figure}[htbp]
    \centering
    \includegraphics[width=6in]{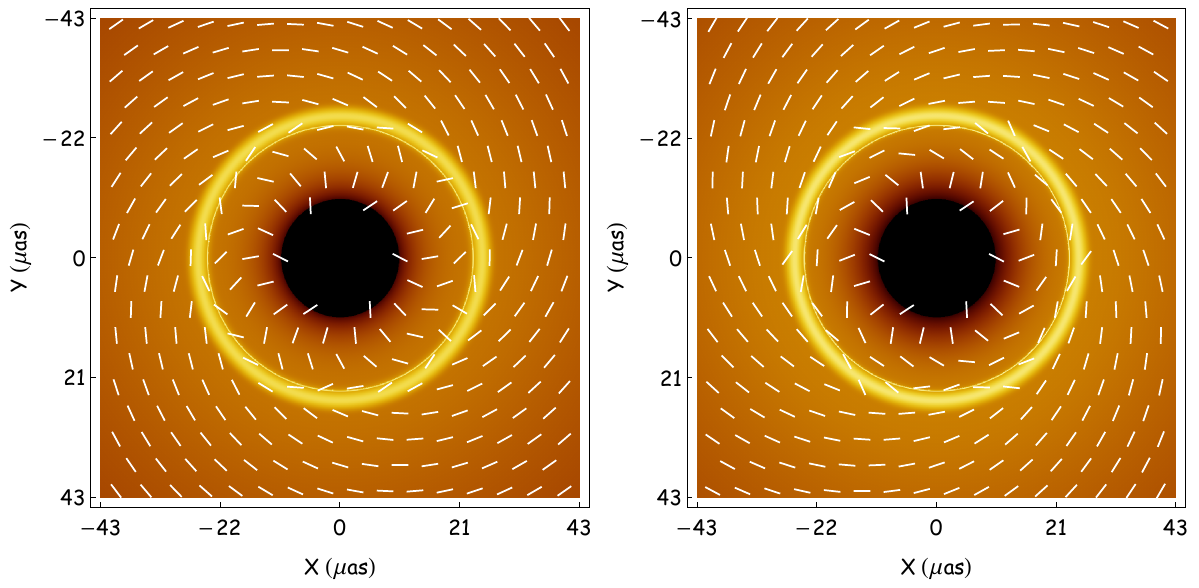}
    \caption{Polarized images of the thin disk with geodesic accretion flow. The left panel corresponds to $L = 1$, aligned with the black hole spin, while the right panel shows $L = -3$, counter-rotating. All other parameters are identical: black hole spin $a = 0.998$, fluid energy parameter $E = 1.2$, and observational angle $\theta_o = 0.1^\circ$.}
    \label{fig:result2}
\end{figure}

 In Fig.~\ref{fig:result2}, we present the polarized images for both initially prograde and retrograde accretion flows. For the initially prograde accretion flow, we set $L = 1$, while for the initially retrograde accretion flow, we choose $L = -3$. The black hole spin parameter and the energy parameter of the particles comprising the accretion flow are fixed at $a = 0.998$ and $E = 1.2$, respectively, with the observation angle set to $0.1^\circ$. The bright ring visible in the figure corresponds to the photon ring, while the short white lines indicate the directions of the polarization vectors at their midpoints. A striking feature is that, outside the photon ring, the spiral orientation formed by the white lines in the two panels of Fig.~\ref{fig:result2} is opposite: in the left panel, the pattern follows a clockwise direction, whereas in the right panel, it follows a counterclockwise one. However, within the photon ring, the spiral orientations of the white lines appear visually indistinguishable between the two panels. Qualitatively, this behaviour remains consistent: for an initially retrograde accretion flow, a transition from retrograde to prograde motion occurs as it approaches sufficiently close to the black hole, leading to a corresponding deflection in the spiral orientation of the polarization lines. In contrast, an initially prograde accretion flow maintains its prograde motion throughout, thereby preserving the spiral orientation of the polarization lines.

\begin{figure}[htbp]
    \centering
    \includegraphics[width=6in]{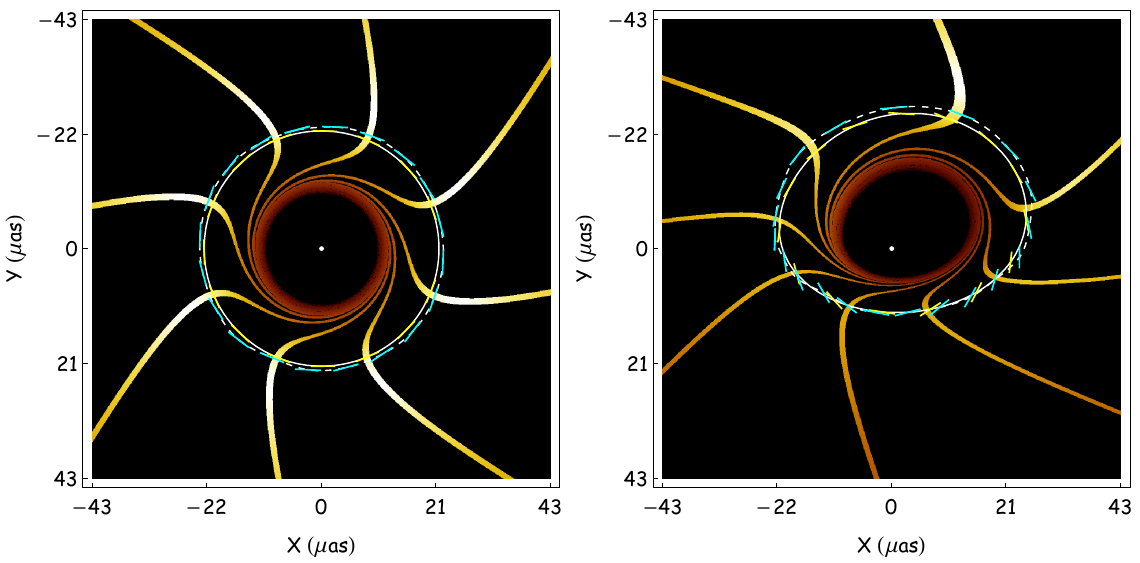}
    \caption{Accretion flow and polarization at specific locations observed from different inclination angles. The dashed contour with short cyan lines indicates the positions where the fluid changes direction on the screen, while the solid contour with short yellow lines  denotes locations where the EVPA is perpendicular to the radial direction on the screen. The white dot represents the center of the screen hereafter. The left panel corresponds to an observation angle of \(0.1^\circ\), while the right panel shows the case for an observation angle of \(50^\circ\). All other parameters are identical: black hole spin \(a=0.998\), \(E=1.2\) and \(L=-3\).}
    \label{fig:result1}
\end{figure}

Although the magnetic field is expected to realign with the turning of streamlines due to flux freezing, their appearances in the image perceived by a distant observer are subject to distinct transformations under the influence of strong gravitational fields. Specifically, gravitational lensing distorts the visual representation of streamlines, while the polarization direction is modified through the parallel transport of bent light rays. As a result, the locations on the image plane where streamlines visibly change direction do not necessarily coincide with those where the polarization orientation shifts.

This disparity is illustrated in Fig.~\ref{fig:result1} through two representative examples. The dashed contour, marked with short cyan lines, traces where the fluid changes direction on the screen, whereas the solid contour with short yellow lines highlights the locations where the EVPA is perpendicular to the radial direction on the image. The white dot denotes the center of the screen, a convention maintained henceforth. The left panel corresponds to an observation angle of $0.1^\circ$, while the right panel depicts the case for a viewing angle of $50^\circ$. All other parameters remain the same: black hole spin $a = 0.998$, energy $E = 1.2$, and angular momentum $L = -3$.

Next, we undertake a detailed analysis of the turning point coordinates on the image—specifically, those of the primary image of the streamline $(\rhot, \vt)$, the polarization flip location $(\rhop, \vp)$, and the true viewpoint of the streamline’s primary image point $(\rhotm, \vtm)$—as well as the relative magnitudes and characteristics of the source positions corresponding to the turning points on the image. 

\subsection{On-axis observers}

Let us first consider a relatively simple scenario, namely that of on-axis observers. It is noteworthy that, for observers situated along the optical axis, the image exhibits central (point) symmetry. Consequently, it is sufficient to confine our analysis to the radial coordinate domain. Noting Eq.~(\ref{rbrr}), we are provided with the expression relating $r_s$ and $\rho$ in Eq. (\ref{rbrr}), it thus suffices to determine the relationship among $\rt$, $\rmt$, and $\rmp$.

\begin{figure}[htbp]
    \centering
    \includegraphics[width=6in]{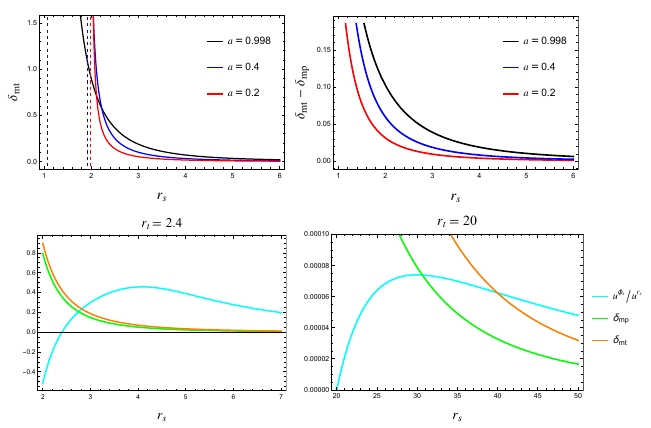}
    \caption{Plots of $\dmt$, $\dmt-\dmp$, and $u^{\phi} / u^{r}$ as functions of the radial coordinate $r_{s}$ of the source.}
    \label{fig:px}
\end{figure}

The critical point $\rhot$, where the image of the accretion flow exhibits a turning behaviour in the $\varphi$-direction, can be identified using the condition $d\varphi = 0$, which leads to  
\bea
\label{phis}
0 = d\phi_s - \frac{2a r_s}{\Delta_s \sqrt{\mathcal{R}_s}}\,dr_s - d\eta\int_{r_s}^{\infty} \frac{a r\, \mathcal{R}^{-3/2}}{\Delta} \frac{\partial \mathcal{R}}{\partial \eta}\,dr = d\phi_s - \frac{2a r_s}{\Delta_s \sqrt{\mathcal{R}_s}}\,dr_s + d\eta \int_{r_s}^{\infty} a r\, \mathcal{R}^{-3/2} dr\,,
\eea
where the subscript “$s$” indicates that the quantities are evaluated at $r = r_s$. When $r_s \gg M$, and using the result derived from Eq.~(\ref{phiaeq}), this equation can be simplified as  
\bea  
d\phi_s = \frac{4a}{r_s^3} \, dr_s.  
\eea  
Dividing both sides of the above equation by $d\tau$, we obtain  
\bea\label{mtaeq}  
\frac{u^{\phi_s}}{u^{r_s}} = \frac{4a}{r_s^3}>0.  
\eea
At the turning point $\rt$ of the accretion flow, the condition $u^{\phi} / u^r = 0$ must be satisfied. Considering that the accretion flow always proceeds inward, i.e., $dr < 0$, the initially retrograde-moving flow satisfies $d\phi < 0$ before reaching the turning point $\rt$, which implies that $u^{\phi} / u^r > 0$. Therefore, from Eq.~(\ref{mtaeq}), it can be inferred that $\rmt > \rt$. However, it is important to note that Eq.~(\ref{mtaeq}) holds only in the regime where $r_s \gg M$. Consequently, for the general case, we must turn to an analysis based on Eq.~(\ref{phis}); That is, if  
\bea \label{mteq} 
d\phi_s = \frac{2a r_s}{\Delta_s \sqrt{\mathcal{R}_s}}\,dr_s - d\eta_s \int_{r_s}^{\infty} \frac{a r}{\mathcal{R}^{3/2}}\,dr > 0,  
\eea  
holds universally, then we may conclude that the inequality $\rmt >\rt$ is likewise always satisfied—an important and noteworthy result. Although we may attempt to derive an expression for $d\phi_s / dr_s$ by relating $d\eta_s$ and $dr_s$ through Eqs. (\ref{reeq}) and (\ref{rbrr}), and subsequently substituting into Eq. (\ref{mteq}), the inherent complexity of the resulting expression renders a general analytical proof of the universal validity of Eq. (\ref{mteq}) rather challenging. Therefore, we proceed with a numerical approach. By selecting various values of the spin parameter $a$, we computed the numerical behaviour of $\dmt\equiv d\phi_s / dr_s$ outside the event horizon and consistently found it to be positive. Several illustrative examples are presented in the left panel of the first row of Fig.~\ref{fig:px}, where the dashed lines indicate the location of the event horizon radius corresponding to each specific spin parameter. At this juncture, we have accumulated ample confidence to assert that the inequality $\rmt > \rt$ holds universally.

It is worth noting that, in the regime where $r_s \gg M$, by combining Eqs. (\ref{mtaeq}) and (\ref{fv}), we arrive at an approximate analytical expression for $\rmt$, expressed as:
\begin{equation}\label{exrmt}
\begin{aligned}
\rmt &= \frac{1}{E^2 - 1} - \frac{2a\left(2\sqrt{E^2 - 1} + E\right)}{L} \\
&= \left[1 + 2\left(1 - \frac{1}{E^2}\right)^{1/2}\right] \rt - 3 - 4\left(1 - \frac{1}{E^2}\right)^{1/2} + \left(1 - \frac{1}{E^2}\right)^{-1}\,.
\end{aligned}
\end{equation}
Furthermore, in the limit where the turning position is sufficiently distant, the ratio $\rmt / \rt$ asymptotically approaches a constant:
\bea
\frac{\rmt}{\rt} = \left[1 + 2\left(1 - \frac{1}{E^2}\right)^{1/2}\right]\,.
\eea
Next, we proceed to further examine the relative magnitudes of $\rmp$, $\rmt$, and $\rt$.  
From Eq.~(\ref{abeq}), by expressing $\mathcal{A}$ and $\mathcal{B}$ in terms of the flow's four-velocity $u^\mu$, one obtains  
\begin{equation}
\begin{aligned}
& \mathcal{A} = \frac{\Phi_0\sqrt{\eta(r_s)}}{u^{r_s}\Delta_s} \left[(r_s^2 + a^2)\Delta_s u^{\phi_s} - a\sqrt{\mathcal{R}_s}\, u^{r_s}\right], \\
& \mathcal{B} = \frac{\Phi_0\sqrt{\eta(r_s)}}{u^{r_s}\Delta_s} \left[a\Delta_s \sqrt{\mathcal{R}_s}\, u^{\phi_s} + \left(\Delta_s \eta(r_s) - 2a^2 r_s\right) u^{r_s}\right],
\end{aligned}
\end{equation}  
where appropriate punctuation and consistent notation are maintained. Recall that the position of the polarity flip $\rhop$ is determined by the condition $f_\mu \cdot \hat{e}_\rho = 0$, which corresponds to $\mathcal{A} = 0$, and leads to a critical relation between the azimuthal and radial components of the flow's velocity:
\bea
\dmp\equiv\frac{d\phi_s}{dr_s} = \frac{a\sqrt{\mathcal{R}_s}}{(r_s^2 + a^2)\Delta_s} > 0\,.
\eea
Given that $\dmp$ is always greater than zero, we can consequently infer that $\rmp$ consistently exceeds $\rt$. Moreover, in the limit where  
$r_s \gg M$, we can similarly obtain an approximate analytical expression for $\rmp$, given by:  
\begin{equation}\label{exrmp}
\begin{aligned}
\rmp &= \frac{1}{E^2 - 1} - \frac{2a\left(\sqrt{E^2 - 1} + E\right)}{L} \\
&= \left[1 + \left(1 - \frac{1}{E^2}\right)^{1/2}\right] \rt - 3 - 2\left(1 - \frac{1}{E^2}\right)^{1/2} + \left(1 - \frac{1}{E^2}\right)^{-1}\,.
\end{aligned}
\end{equation}
Also, in the limit where the turning position is sufficiently distant, the ratio $\rmp / \rt$ asymptotically approaches a constant:  
\begin{equation}
\frac{\rmp}{\rt} = \left[1 + \left(1 - \frac{1}{E^2}\right)^{1/2}\right]\,.
\end{equation}
Furthermore, by comparing Eqs.~(\ref{exrmt}) and (\ref{exrmp}), we arrive at  
\begin{equation}
\rmt - \rmp = \left(1 - \frac{1}{E^2}\right)^{1/2} \left(\rt - 2\right) > 0\,,
\end{equation}  
which means $\rmt > \rmp$ when $\rt \gg M$. However, this alone does not suffice to demonstrate that the inequality holds for a general $\rt$. Given the analytical difficulty in proving that $\rmt > \rmp$ always holds, we resort to numerical verification. First, we numerically compare $\dmt - \dmp$ under various spin parameters and find that $\dmt - \dmp$ consistently remains positive, as illustrated in the right panel of the first row in Fig.~\ref{fig:px}. In addition, we numerically obtain the profile of $u^\phi / u^r$ for an accretion flow moving along a geodesic. Although $u^\phi / u^r$ is not a monotonic function of $r_s$ and exhibits a local maximum, it is important to note that there exists only a single intersection point between $u^\phi / u^r$ and each of $\dmt$ and $\dmp$, respectively. Since the curve of $\dmt$ consistently lies above that of $\dmp$, it follows that as $r_s$ increases, $u^\phi / u^r$ must first intersect $\dmp$ and then $\dmt$. Consequently, the corresponding relation $\rmp < \rmt$ is always satisfied, as shown in the two panels of the second row in Fig.~\ref{fig:px}.

In conclusion, for on-axis observers and accretion flows moving along geodesics, we find that the inequality  
$\rt < \rmp < \rmt$ consistently holds. Combined with Eq.~(\ref{rbrr}), this implies that  
$\rhotm < \rhop < \rhot$ is always satisfied—one of the key findings of our present work.

\begin{figure}[htbp]
    \centering
    \includegraphics[width=5in]{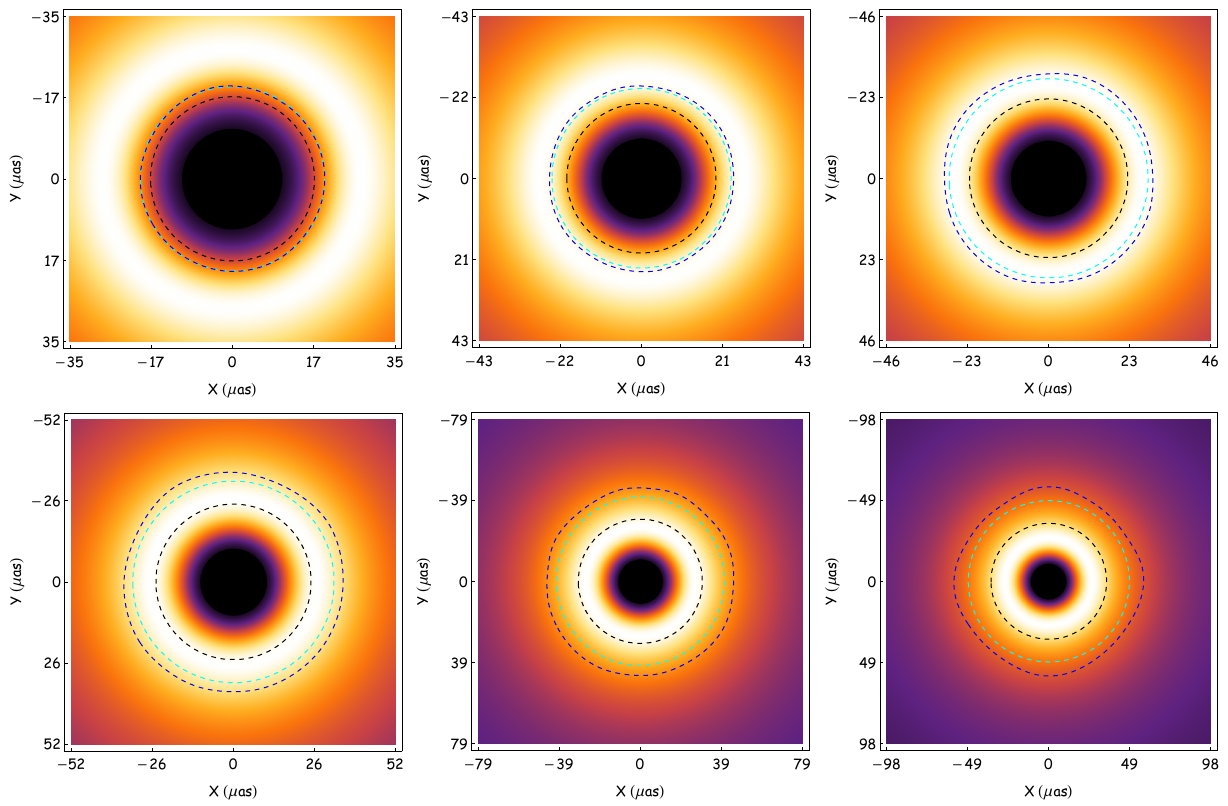}
    \caption{Primary images of accretion disks, characterized by retrograde geodesic flows with various values of \(\rt\). From top to bottom and left to right, \(\rt=2.5\,,3\,,3.5\,,4\,,5\,,6\), respectively. The dashed contours in each panel, from inner to outer, correspond to \(\rhotm\) (black), \(\rhop\) (cyan) and \(\rhot\) (blue). We set \(\theta_o=0.1^\circ\,,a=0.998\).}
    \label{fig:kerr1}
\end{figure}

To further substantiate the aforementioned conclusions, we directly determine the magnitudes of $\rhotm$, $\rhop$, and $\rhot$ from the black hole images. The corresponding results are displayed in Fig.~\ref{fig:kerr1}. As illustrated in the figure, we present direct images of retrograde geodesic accretion flows for various values of $\rt$, demonstrating the radial dependence of the frame-dragging effect. The panels are arranged sequentially with $\rt = 2.5$, 3, 3.5, 4, 5, 6, from top to bottom and left to right. The observer’s inclination angle is fixed at $\theta_o = 0.1^\circ$. Within each panel, the dashed contours—ordered from innermost to outermost—represent $\rhotm$ (black), $\rhop$ (cyan), and $\rhot$ (blue), respectively. This convention is maintained throughout the manuscript. For improved visual clarity, distinct field-of-view angles have been employed in the numerical simulations. Taken as a whole, we observe that the inequality $\rhotm < \rhop < \rhot$ consistently holds. Furthermore, several nuanced features are evident: the separation between $\rhot$ and $\rhop$ is markedly smaller than the individual offsets of each from $\rhotm$. As $\rt$ increases, the separations between $\rhotm$ and $\rhot$, as well as between $\rhotm$ and $\rhop$, become progressively more pronounced. In the top-left panel, corresponding to $\rt = 2.5$, the cyan and blue contours almost overlap, whereas in the subsequent panels, the divergence between them becomes increasingly significant.

\subsection{Off-axis observers}

We now consider a more general scenario involving an off-axis observer. Unlike the on-axis case, the image observed by an off-axis observer no longer exhibits central (point) symmetry. As a result, both $\rho$ and $\varphi$ become nontrivial variables. Consequently, in addition to $\rho$, one must also take into account the dependence on $\varphi$. The semi-analytic approach employed in the previous subsection thus becomes inapplicable. For this reason, we adopt a numerical ray-tracing method \cite{Hu:2020usx, Zhong:2021mty} to analyze the case of an off-axis observer.

\begin{figure}[htbp]
    \centering
    \includegraphics[width=5in]{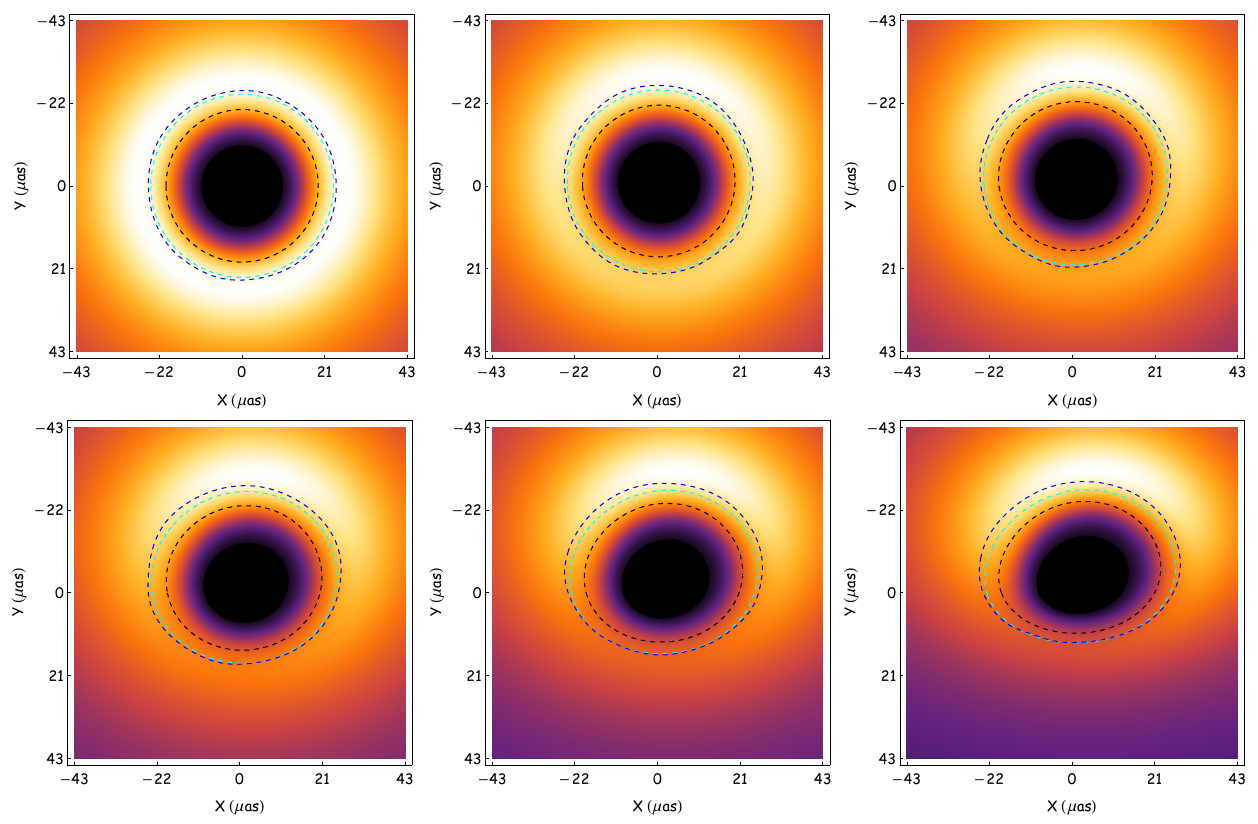}
    \caption{Primary images of accretion disks with retrograde geodesic flows for various observer inclinations \(\theta_o\). From top to bottom and left to right, \(\theta_o=0.1^\circ\,,10^\circ\,,20^\circ\,,30^\circ\,,40^\circ\,,50^\circ\), respectively. The dashed contours in each panel, from inner to outer, correspond to \(\rhotm\) (black), \(\rhop\) (cyan) and \(\rhot\) (blue). We fix \(\rt=3\,,a=0.998\).}
    \label{fig:kerr2}
\end{figure}

To further examine the observer-dependent features of frame dragging, we display in Fig.~\ref{fig:kerr2} the primary images of retrograde geodesic accretion flows for a fixed turning point radius $\rt$, viewed at various inclination angles $\theta_o$. The panels are arranged from left to right and top to bottom, corresponding to $\theta_o = 0.1^\circ$, $10^\circ$, $20^\circ$, $30^\circ$, $40^\circ$, and $50^\circ$, respectively. Several key features emerge from these off-axis images. First, for a fixed azimuthal angle $\varphi$, the inequality $\rhotm < \rhop < \rhot$ persists across all inclination angles, consistent with the on-axis case. Moreover, the separation between $\rhot$ and $\rhop$ remains significantly smaller than the individual displacements of either from $\rhotm$. As the inclination angle increases, the loss of central symmetry becomes evident, and the contours undergo progressively stronger deformations. While the qualitative trends in distortion remain similar among the three contours, their degree of asymmetry exhibits a pronounced azimuthal dependence at larger inclination angles. For near-axis observers, the contour separations remain nearly uniform in $\varphi$. By contrast, in the bottom-right panel corresponding to $\theta_o = 50^\circ$, the contours—$\rhotm$, $\rhop$, and $\rhot$—are clearly separated in the upper part of the image, while in the lower part, the separation diminishes substantially, with $\rhot$ and $\rhop$ nearly coinciding.

Next, to further quantify the differences among these three contours, we define the enclosed area of each as  
\bea
S_i = \frac{1}{2} \int_{0}^{2\pi} \left[ \rho_i(\varphi) \right]^2\, d\varphi = \pi \bar{\rho}_i^2,  
\eea  
where the subscript $i \in \{\text{tm}, \text{p}, \text{t}\}$ refers to the regions enclosed by $\rhotm$, $\rhop$, and $\rhot$, respectively, and $\bar{\rho}_i$ represents the area-averaged radius of the corresponding contour, i.e., the radius of a circle having the same enclosed area $S_i$. We then introduce the area ratios $S_\text{tm} / S_\text{p}$ and $S_\text{tm} / S_\text{t}$ as quantitative measures of their mutual deviations, and denote a generic area ratio by $\epsilon$.

\begin{figure}[htbp]
    \centering
    \includegraphics[width=6.5in]{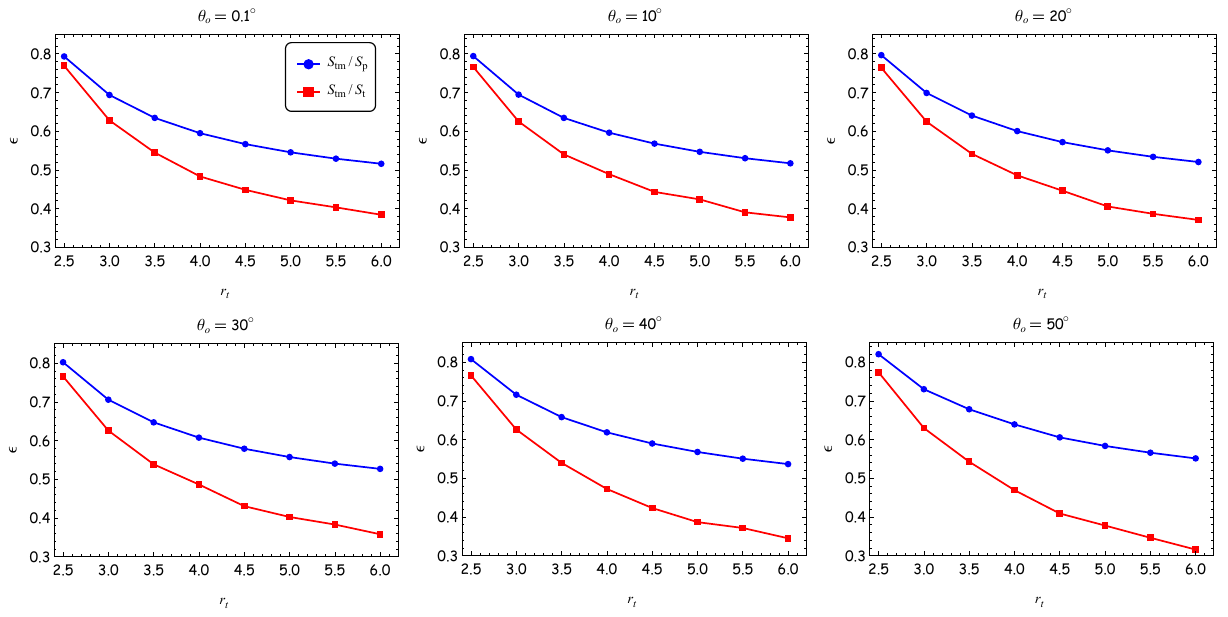}
    \caption{Graphical representations of the function \(\epsilon(\rt)\) for retrograde geodesic flows at various observer inclinations.}
    \label{fig:kerr3}
\end{figure}

In Fig.~\ref{fig:kerr3}, we display the dependence of \(\epsilon\) on \(\rt\) for inclination angles  \(\theta_o=0.1^\circ\,,10^\circ\,,20^\circ\,,30^\circ\,,40^\circ\,,50^\circ\), arranged from top to bottom and left to right. In each subplot (and in all subsequent figures), the blue line denotes the area ratio \(S_\text{tm}/S_\text{p}\), while the red line corresponds to \(S_\text{tm}/S_\text{t}\). For a fixed inclination (i.e., within each subplot), both ratios decrease markedly as \(\rt\) increases, and the separation between them widens with larger \(\rt\). Moreover, as \(\theta_o\) increases, the disparity between \(S_\text{tm}/S_\text{p}\) and \(S_\text{tm}/S_\text{t}\) becomes progressively more pronounced. This indicates that the average radii satisfy $\bar{\rho}_\text{tm}<\bar{\rho}_\text{p}<\bar{\rho}_\text{t}$, and that the differences among these three average radii increases as $\rt$ increases. In addition, the difference between $\bar{\rho}_\text{p}$ and $\bar{\rho}_\text{t}$ increases steadily with increasing inclination angle $\theta_o$.

\begin{figure}[htbp]
    \centering
    \includegraphics[width=6.5in]{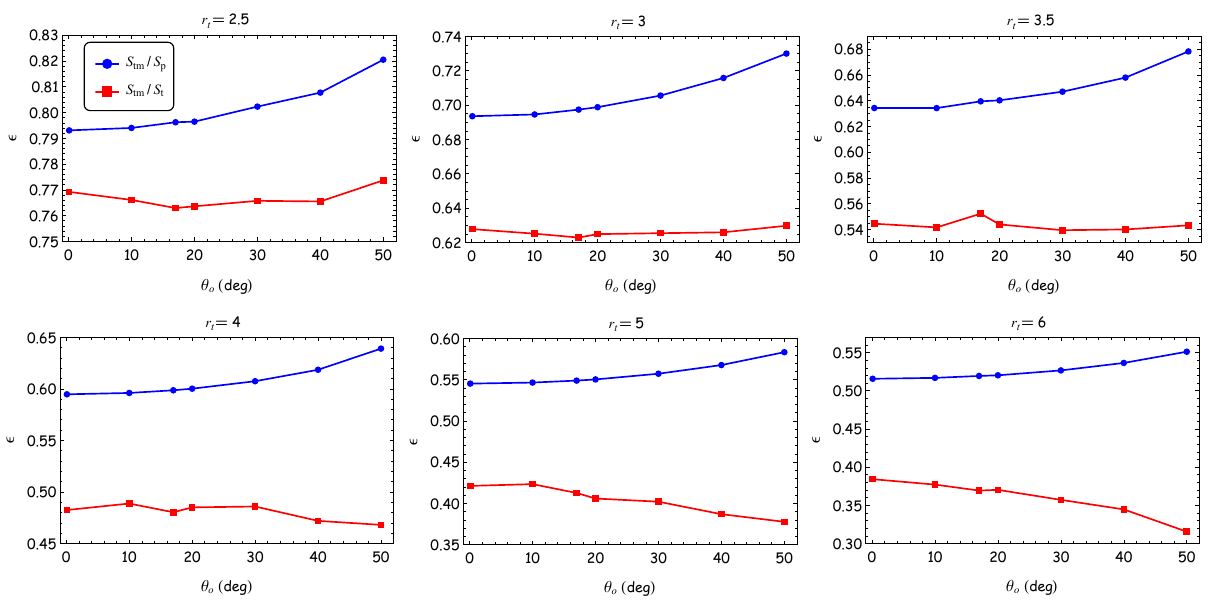}
    \caption{Graphical representations of the function \(\epsilon(\theta_o)\) for retrograde geodesic flows at various fixed values of \(\rt\).}
    \label{fig:kerr4}
\end{figure}

In Fig.~\ref{fig:kerr4}, we present the evolution of the area ratios \(S_\text{tm}/S_\text{p}\) and \(S_\text{tm}/S_\text{t}\) as functions of the inclination angle \(\theta_o\), for fixed values of \(\rt\). We observe that \(S_\text{tm}/S_\text{p}\)  increases monotonically with $\theta_o$ across all values of $\rt$. In contrast, the behavior of $S_\text{tm}/S_\text{t}$ is more nuanced: At small values of  $\rt$, the ratio  $S_\text{tm}/S_\text{t}$ increases mildly with $\theta_o$, but with a smaller amplitude compared to $S_\text{tm}/S_\text{p}$. Conversely, at large $\rt$, $S_\text{tm}/S_\text{t}$ decreases rapidly as $\theta_o$ increases, with a steeper gradient than $S_\text{tm}/S_\text{p}$. Thus, the separation between the two ratios becomes more prominent with increasing \(\theta_o\), highlighting the growing impact of inclination. These trends are also reflected in the average radii $\bar{\rho}_\text{tm}$,  $\bar{\rho}_\text{p}$ and $\bar{\rho}_\text{t}$, where the differences between $\bar{\rho}_\text{p}$ and $\bar{\rho}_\text{tm}$, and between $\bar{\rho}_\text{t}$ and $\bar{\rho}_\text{p}$, increase with the inclination angle. Moreover, the difference between $\bar{\rho}_\text{tm}$ and $\bar{\rho}_\text{t}$ becomes more pronounced at larger values of $\rt$, In particular, while the difference between $\bar{\rho}_\text{tm}$ and $\bar{\rho}_\text{t}$ decreases as the inclination angle decreases in regions closer to the black hole, it increases significantly at greater distances from the black hole. This behavior illustrates the interplay between inclination-dependent asymmetry and the spatial separation of the contours. In particular, they highlight the growing anisotropy in the image structure as one moves away from the spin axis. Together with Fig.~\ref{fig:kerr3}, this set of results provides a more complete quantitative picture of the geometric deformation induced by frame dragging.

\section{Summary}\label{sec4}

We investigated the polarization of black hole images due to frame dragging in the Kerr spacetime. The source was modeled as a thin disk composed of initially retrograde timelike plunging geodesics, with a magnetic field aligned with the streamlines, and synchrotron emission as the radiation mechanism. In Boyer–Lindquist coordinates, we analytically determined the turning point of the fluid motion, which coincided with the flip of the polarization polarity.

We examined the relation between the turning points of the accretion flow $(\rt, \pt)$, primary image’s turning point $(\rmt, \pmt)$, and the primary image’s EVPA flip point $(\rmp, \pmp)$ (noted that, the corresponding coordinates on the image plane were denoted as $(\rhotm, \vtm)$, $(\rhot, \vt)$, $(\rhop, \vp)$). We showed both semi-analytically and numerically that for on-axis observers, $\rt<\rmp < \rmt$ and $\rhotm < \rhop < \rhot$. and in the limit $r_s\gg M$ we obtained approximate analytic expressions for $\rmt$ and $\rmp$. We found that when $\rt$ was sufficiently large, both $\rmt/\rt$ and $\rmp/\rt$ asymptotically approached constants that depended only on the flow energy. By employing both numerical and analytical methods, we performed a cross-validation of our results and found them to be consistent. For off-axis observers, we numerically analyzed the effect of the  inclination angle and turning point $\rt$ on these three characteristic locations and introduced a dimensionless area ratio  $\epsilon$ to quantify the relative size of the region enclosed by them.

Our results showed that, for any inclination angle, the ordering $\rhotm < \rhop < \rhot$ always held for the same azimuthal angle, and the separation between $\rhop$ and $\rhot$ was much smaller than the separations between either of them and $\rhotm$. Furthermore, the separations among $\rhotm$, $\rhop$ and $\rhot$ exhibited a non-uniform dependence on the azimuthal angle. To quantify these features, we introduced the enclosed area and the area-averaged radius of the regions bounded by $(\rhotm, \vtm)$, $(\rhop, \vp)$ and $(\rhot, \vt)$. As $\rt$ increased, the separations among the three averaged radius became larger. As $\theta_o$ increased, the relative separations between $\bar{\rho}_\text{tm}$ and $\bar{\rho}_\text{p}$, and between $\bar{\rho}_\text{tm}$ and $\bar{\rho}_\text{t}$, depended on the distance from the black hole, whereas the gap between  $\bar{\rho}_\text{p}$ and $\bar{\rho}_\text{t}$ increased monotonically.

We conclude this paper with several prospects for future exploration. Due to the complexity of the full expressions for the critical points in the case of on-axis observers, our analytical treatment necessarily employed certain approximations, and a complete closed-form solution has yet to be derived. Furthermore, in this work we considered only geodesic flows; extending the analysis to more general accretion models would be of significant interest.

\section*{Acknowledgments}
We are grateful to Yehui Hou and Zhenyu Zhang for insightful discussions. The work is partly supported by NSFC Grant No. 12275004 and 12205013. M. Guo is also supported by Open Fund of Key Laboratory of Multiscale Spin Physics (Ministry of Education), Beijing Normal University.

\appendix
\label{app}

\bibliographystyle{utphys}
\bibliography{reference}

\end{document}